\documentclass[twocolumn]{aastex63}
\usepackage{lineno}
\usepackage{soul}
\received{Month day, 2019}
\revised{Month day, 2019}
\accepted{Month day, 2019}
\submitjournal{ApJ}

\shorttitle{GASTRO II}
\shortauthors{Amarante et al.}
\graphicspath{{./}{figures/}}

\begin{document}

\title{{\tt GASTRO} library II: Exploring Chemical Bimodalities in Disk Galaxies with GSE-like Mergers and Massive Star-forming Clumps }


\correspondingauthor{ \newline Jo\~ao A. S. Amarante (joaoant@gmail.com),\newline Zhao-Yu Li (lizy.astro@sjtu.edu.cn) }

\author[0000-0002-7662-5475]{Jo\~ao A. S. Amarante}%
\affil{Department of Astronomy, School of Physics and Astronomy, \\ Shanghai Jiao Tong University, 800 Dongchuan Road, Shanghai, 200240, China}
\affil{State Key Laboratory of Dark Matter Physics, School of Physics and Astronomy,\\ Shanghai Jiao Tong University, Shanghai, 200240, China}
\affiliation{Institut de Ciencies del Cosmos (ICCUB), Universitat de Barcelona (IEEC-UB), Martí i Franquès 1, E-08028 Barcelona, Spain}
\author[0000-0003-3922-7336]{Chervin F. P. Laporte}
\affiliation{LIRA, Observatoire de Paris, Universit\'e PSL, Sorbonne Universit\'e, Universit\'e Paris Cit\'e, CY Cergy Paris Universit\'e, CNRS, 92190 Meudon, France}
\affiliation{Institut de Ciencies del Cosmos (ICCUB), Universitat de Barcelona (IEEC-UB), Martí i Franquès 1, E-08028 Barcelona, Spain}
\affiliation{Kavli IPMU (WPI), UTIAS, The University of Tokyo, Kashiwa, Chiba 277-8583, Japan}
\author[0000-0001-7902-0116]{Victor P. Debattista}
\affil{Jeremiah Horrocks Institute, University of Lancashire, Preston, PR1 2HE, UK}
\author[0000-0002-0740-1507]{Leandro {Beraldo e Silva}}
\affil{Observat\'{o}rio Nacional, Rua General Jos\'{e} Cristino 77, Bairro S\~ao Crist\'ov\~ao, Rio de Janeiro, 20921-400, RJ, Brasil}
\author[0000-0002-9269-8287]{Guilherme~Limberg}
\affiliation{Kavli Institute for Cosmological Physics, University of Chicago, 5640 S. Ellis Avenue, Chicago, IL 60637, USA}
\affiliation{Department of Astronomy \& Astrophysics, University of Chicago, 5640 S. Ellis Avenue, Chicago, IL 60637, USA}
\author[0000-0002-0537-4146]{H\'elio D. Perottoni}
\affil{Observat\'{o}rio Nacional, Rua General Jos\'{e} Cristino 77, Bairro S\~ao Crist\'ov\~ao, Rio de Janeiro, 20921-400, RJ, Brasil}
\author[0000-0001-5017-7021]{Zhao-Yu Li}
\affil{Department of Astronomy, School of Physics and Astronomy, \\ Shanghai Jiao Tong University, 800 Dongchuan Road, Shanghai, 200240, China}
\affil{State Key Laboratory of Dark Matter Physics, School of Physics and Astronomy,\\ Shanghai Jiao Tong University, Shanghai, 200240, China}
\author[0000-0003-3382-1051]{Lais Borbolato}
\affil{Universidade de S\~ao Paulo, Instituto de Astronomia, Geof\'isica e Ci\^encias Atmosf\'ericas, Departamento de Astronomia, \\ SP 05508-090, S\~ao Paulo, Brasil}
\author[0000-0001-7902-0116]{Karl Fiteni}
\affil{Como Lake centre for AstroPhysics (CLAP), DiSAT, Università dell’Insubria, Via Valleggio 11, 22100 Como, Italy}
\affil{Institute of Space Sciences \& Astronomy, University of Malta, Msida MSD 2080, Malta}
\author[0000-0001-6655-854X]{Chengye Cao}
\affil{Department of Astronomy, School of Physics and Astronomy, \\ Shanghai Jiao Tong University, 800 Dongchuan Road, Shanghai, 200240, China}
\affil{State Key Laboratory of Dark Matter Physics, School of Physics and Astronomy,\\ Shanghai Jiao Tong University, Shanghai, 200240, China}
\author[0000-0003-3523-7633]{Nathan Deg}
\affil{Department of Physics, Engineering Physics, and Astronomy, Queen's University, Kingston ON K7L~3N6, Canada}
\author[0000-0002-3343-6615]{Tigran Khachaturyants}%
\affil{Department of Astronomy, School of Physics and Astronomy, \\ Shanghai Jiao Tong University, 800 Dongchuan Road, Shanghai, 200240, China}
\affil{State Key Laboratory of Dark Matter Physics, School of Physics and Astronomy,\\ Shanghai Jiao Tong University, Shanghai, 200240, China}
\author[0009-0003-8415-556X]{Xiaojie Liao}%
\affil{Department of Astronomy, School of Physics and Astronomy, \\ Shanghai Jiao Tong University, 800 Dongchuan Road, Shanghai, 200240, China}
\affil{State Key Laboratory of Dark Matter Physics, School of Physics and Astronomy,\\ Shanghai Jiao Tong University, Shanghai, 200240, China}
%




\begin{abstract}
We use several smoothed particle hydrodynamics+N-body models as part of the {\tt GASTRO} library to study the role of high-density star-forming clumpy regions and a single merger on the formation of the $\alpha$-rich and $\alpha$-poor populations in the disk galaxies. These experiments are tailored to mimic what is expected to be the Gaia-Sausage/Enceladus (GSE) accretion event, which occurred circa 10 Gyr ago in the Milky Way (MW). We find that either an early clumpy phase or a retrograde merger significantly reduces the star formation rate (SFR) of the disk, giving rise to a chemical bimodality qualitatively similar to the MW's. The decrease of the SFR as the cause of the chemical bimodality is consistent with previous idealized and cosmological simulations. On the other hand, a prograde radial merger does not significantly modify the SFR of the disk,  resulting in no clear chemical bimodality. We further show that stars originating from the inner regions ($R_{form}<4$ kpc) do not create the disk's chemical bimodality, although they can enhance it. Finally, only the models with an early clumpy phase can produce a significant fraction of old, age $>11$ Gyr, $\alpha-$poor stars with disk-like orbits, similar to what has been recently observed in the MW. Our results strengthen the case of clumpy disky galaxies observed at redshift $z\approx 1-2$ as likely progenitors of our Galaxy.
\end{abstract}



\section{Introduction}\label{sec:intro}
In the current cosmological $\Lambda$-Cold Dark Matter model \citep{planck2020}, galaxies grow hierarchically \citep{searle-zinn1978,efstathiou1985} through the condensation of gas at the centre of dark matter (DM) halos \citep{white1978} and forming stars, the entire process being regulated by feedback mechanisms \citep[e.g.][]{dekel1986, white1991}. As such, the Milky Way (MW), being a disk galaxy, must have had a fairly quiescent merger history \citep[e.g.][]{freeman2002, bland-hawthorn16} in order to form and maintain its disk morphology over several Gyr \citep[see e.g.][]{xiang2025}.

When decomposed in terms of chemistry, the Galaxy appears, to first order, to be composed of two disk chemical sequences: an $\alpha-$rich disk with large scale height and short scale length, and an $\alpha-$poor sequence with a short scale height and larger scale length \citep{bovy+2012}. Dissections of the disk as a function of height and radius \citep[e.g.][]{hayden2015, queiroz2020} reveal a number of other subtleties pointing to potentially different chemical dichotomies with separate origins. Particularly, in the inner region of the MW, at galactocentric radius $R_{gc}<4$ kpc, the chemical sequence appears as a single track with a high$-\alpha$ disk and a metal-rich $\alpha-$poor bimodal distribution separated by a dearth of stars between them \citep{queiroz+2021}.
This has been suggested as the outcome of an episode of significantly reduced star formation in the central region of the Galaxy, potentially caused by the bar, a merger-induced starburst, or AGN activity \citep[see e.g.][]{haywood24, grand2020, beane25,beane25bagn, orkney+2025}. Alternatively, \citet{debattista+2023} found that an increased star formation caused by massive clumps reproduces the MW's trend. \par
In the solar neighborhood, on the other hand, the chemical dichotomy takes a different form with the $\alpha-$poor sequence extending to the metal-poor end of the high$-\alpha$ sequence. This has been suggested as evidence for inflow of pristine gas from the outer region of the Milky Way \citep[e.g.][]{haywood2013}. But see \citet{khoperskov+2025paperII} who recently found a less pronounced metal-poor tail for the low$-\alpha$ disk based on an orbit-superposition approach \citep{khoperspov+2025paperI}. Both idealized and cosmological simulations find that this portion of the chemical bi-modality is typically formed through gas condensation from the hot-circumgalactic medium halo \citep[e.g.][]{Khoperskov2021, renaud+2021, parul2025, orkney+2025}.\par
A generic result across all cosmological simulations is that the formation of the high$-$ and low$-\alpha$ sequences is sequential (e.g. \citealt{orkney+2025}, but see also, e.g. \citealt{agertz2021,renaud+2021} for a coeval formation scenario) with the high$-\alpha$ sequence forming at high redshift inheriting its high velocity dispersion from the more turbulent state of the interstellar medium (ISM) and the low$-\alpha$ sequence forming during the more quiescent phase of galaxy growth after $z\sim2$ and inheriting the lower velocity dispersion for its stars at birth \citep{bird13, minchev15, navarro18}. Whether a galaxy develops two chemical sequences or just a single track seems to also depend mostly on the gas reservoir of the galaxy and star formation history \citep{orkney+2025}, as such chemical bimodalities may not be a generic feature of every galaxy, with M31 being an example of a disk galaxy with currently no apparent evidence for a chemical bimodality according to the latest JWST data (\citealt{nidever2024}, but see \citealt{kobayashi+2023}). \par
Old low$-\alpha$ disk stars have been observed in the solar neighborhood \citep{silva-aguirre+2018} and in the outer edge of the Galaxy \citep{laporte+2020}. In recent years, there has been growing evidence that these stars make up a significant population of the early MW disk \citep{beraldo+2021, gent+2024, nepal+2024, borbolato+2025} and thus suggest an almost co-evolutionary formation for the high and low$-\alpha$ sequence.
Observations of high-redshift galaxies often show evidence for massive star-forming clumps \citep[e.g.][]{elmegreen2005, guo2018, sok+2025}. Idealized simulations of hot halos cooling within DM halos find that these clumps, although having life times of order of only a few hundred Myr \citep{garver+2023}, can self-enrich and also dynamically heat their surroundings, such as to produce the high$-\alpha$ sequence, while the low$-\alpha$ sequence starts forming more extended, out of the clumps, and with a lower SFR \citep{clarke+2019, beraldo+2021}. \par
Moreover, our Galaxy is known to have undergone a significant merger around $z\sim2$ with a galaxy dubbed Gaia-Sausage-Enceladus \citep[GSE,][]{belo2018, helmi2018, haywood2018} which may have triggered subsequent star formation \citep[e.g.][]{ bignone+2019,grand2020, ciuca+2022} and/or excited the formation of the low$-\alpha$ sequence through inflow of metal-poor gas from the surrounding circumgalactic medium in the outer disk of the galaxy \citep{renaud+2021, orkney+2025}). \par
The initial experiments of \cite{clarke+2019} considered the evolution of a MW-like halo in isolation, varying only the efficiency of supernova feedback, which resulted in an evolution with a clumpy and non-clumpy mode of star formation at low feedback efficiency, and only a non-clumpy mode at high efficiency. They found that in the absence of clumpy star formation, no chemical bimodality forms. However, given the unambiguous evidence for the GSE merger \citep{chiba, brook2004, meza2005,nissen-schuster2010, leaman+2013,deason+2013,belo2018, helmi2018, haywood2018,myeong+2018, limberg+2022}, one would also need to consider its impact on the later formation and evolution of the MW and its chemical sequences, under scenarios with and without clumpy star formation. This is the goal of this contribution, which aims to study the formation of chemical bimodalities and differences between different star formation models in the presence of a  GSE merger through a library of hydrodynamical simulations: the {\tt GASTRO} library.\par
This work is organized as follows: in Section \ref{sec:models} we present the {\tt GASTRO} library and discuss the evolution of the host and GSE-like galaxy for each model considered in Section \ref{sec:modelevol}. In Section \ref{sec:chembi}, we contrast the various chemical sequences resulting from each experiment and how the different chemical dichotomies arise. We then revisit when the $\alpha-$poor sequence starts to dominate the disk in numerical experiments in the presence of clumpy and non-clumpy star formation with a GSE-like merger in Section \ref{sec:age}. In Section \ref{sec:old}, we focus our attention on the emergence of old (age greater 10 Gyr), metal-poor low$-\alpha$ disk stars for each model. Finally,  we discuss the implications of our results in Section \ref{sec:discussion} and present our conclusions in Section \ref{sec:conclusion}.

\section{{\tt GASTRO} library}\label{sec:models}
As introduced in the previous section, isolated disk-galaxy models with an early clumpy phase reproduce several key aspects of the MW: the chemical bi-modality in the disk \citep{clarke+2019}, the chemo-geometric correlations of $\alpha-$poor/$\alpha-$rich and thin/thick disks \citep{beraldo+2020}, the heated thick disk \citep{amarante+2020splash}, the old, age $> 10$ Gyr, $\alpha-$poor population \citep{beraldo+2021}, and the bulge's chemistry \citep{debattista+2023}. Motivated by these, we developed the \textit{Gaia-EncelAdus-Sausage Timing, chemistRy and Orbit} library, {\tt GASTRO}\footnote{More information and updates about {\tt GASTRO}, can be found at \url{https://jasamarante.github.io/gastro}}\citep{amarante+2022}, which includes a single merger event mimicking the GSE dwarf merger of the MW \citep{belo2018,helmi2018, haywood2018}. {\tt GASTRO}  comprises a number of N-body + Smoothed Particle Hydrodynamics (SPH) simulations, evolved with {\tt GASOLINE} \citep{wadsley}, exploring the effects of clumps and mergers in a disk-like galaxy. A subset of these realizations was used to study the chemodynamical signatures of a GSE-like merger in the MW's halo, allowing us to study the rich chemodynamical space created by the GSE debris and show that some of the known substructures of the MW stellar halo may share a common origin with it \citep{amarante+2022}. \par
\subsection{Input physics}\label{sec:gridphyics}
We refer to \citet{amarante+2022} for a detailed description of the input physics in the galaxy formation model used to run the simulations. The current work explores models that couple either 20\% or 80\% of the $10^{51}$ erg per supernova as thermal energy injected into the ISM following the blastwave implementation \citep{stinson}. In these tailored numerical experiments, the lower feedback regime prevents the gas from being expelled from the clumps, thus allowing a clumpy star formation mode.
Hereafter, we will call the simulations with 20\% and 80\% feedback as clumpy and non-clumpy, respectively.
\subsection{Galaxies initial condition}
The initial conditions (ICs) for the MW-like galaxy consist of a dark matter (DM) halo following a Navarro-Frenk-White (NFW, \citealt{nfw}) density profile with virial radius $r_{200}\approx 200$ kpc, mass $10^{12}\, {\rm M_{\odot}}$, and scale radius 30 kpc. There are a total of $10^6$ DM particles with two distinct masses: $10^6\,{\rm M_{\odot}}$ and $3.5\times10^6\,{\rm M_{\odot}}$ inside and outside 200 kpc, respectively. The gas corona has a total mass of $1.4\times10^{11}\,{\rm M_{\odot}}$ and follows the same radial profile as the DM halo. There are $10^6$ gas particles, each starting with an initial mass of $1.4\times10^5\,{\rm M_{\odot}}$. The spin parameter of the gas is set to $\lambda=0.065$ \citep{bullock+2001}, so that as it cools, via metal-line cooling \citep{shen2010}, it settles into a disk. The softening length for the gas and DM particles is 50 pc and 100 pc, respectively. Star particles inherit the softening length and chemistry from their parent gas particle. They form with probability 5\% directly from the gas particles whenever the temperature drops below 15,000 K and the density exceeds $1\, {\rm cm^{-3}}$ in a convergent flow \citep{stinson}. \par
For the merging satellite galaxy, we use the dwarf galaxy model D1 described in \citet{amarante+2022}. These ICs are set up using {\tt GalacticICS} \citep{kuijken-dubinski1995, widrow-dubinsky2005, widrow+2008, deg+2019}. The dwarf has a total of $10^5$ and $2\times 10^4$ DM and gas particles, respectively. The dwarf galaxy DM halo is parametrized with an NFW density profile with a virial radius of $\approx91$ kpc. We set the truncation radius to 50 kpc with a truncation width of 5 kpc (see \citealt{widrow+2008} for details on the truncation method).
The gas disk is built exponentially, with an initial kinematic temperature of 40,000 K (this temperature provides the pressure to keep the gas disk in hydrostatic equilibrium) and a scale radius of 4 kpc. Its scale length and initial total mass are 1 kpc and $1.4\times10^9\,{\rm M_{\odot}}$, respectively. 
\subsection{Dwarf initial orbit setup}
For all the models, the dwarf galaxy is set at a starting distance of $r_{gc} = 200$ kpc from the MW host. To understand the effect of the orbit impact parameter on the evolution of an MW-like galaxy disk, we focus on a set of models that includes both prograde and retrograde mergers of different orbit circularities. A merger is retrograde (prograde) when its initial orbital angular momentum, $L_{z,0}$, is negative (positive) with respect to the MW-like galaxy stellar vertical angular momentum, $L_z$, component. We consider mo\-dels with orbit circularity, $\eta\equiv L_{z,0}/L_c(E)$, set to 0.3 and 0.5, where $L_c(E)$ is the angular momentum of a circular orbit with energy $E$ in the disk plane. All of the models have their orbit inclination relative to the host galaxy's disk set to $\theta=15^{\circ}$. Our orbital setups were motivated by \citet{naidu2021}, who used N-body-only models to explore a grid of orbital parameters of a GSE-like merger that best reproduces the present-day MW's stellar halo kinematics.
\section{Model evolution}\label{sec:modelevol}%
\begin{table*}
 \begin{center}
 \begin{tabular}{cccccccc}
 \hline
 Model & Clumpy & Type of merger & $\eta$ & Chemical bimodality  & $h_R$ [kpc] & $h_{z1}$ [kpc] &  $h_{z2}$ [kpc] \\
  \hline
  c.iso & Yes & no merger & -- & Yes & 2.04 & 0.24 & 0.97\\
  nc.iso & No & no merger & -- & No  & 2.5 & 0.33 & --\\
  c.p.c03 & Yes & prograde  &  0.3 & ${\rm Yes^*}$ & 3.7 & 0.33 & 1.14\\ 
  c.p.c05 & Yes & prograde  &  0.5 & ${\rm Yes^*}$  & 4.11 & 0.38 & 1.41\\
  c.r.c03 & Yes & retrograde  &  0.3 & Yes  & 1.35 & 0.2 & 0.87\\
  c.r.c05 & Yes & retrograde  &  0.5 & Yes  & 1.41 & 0.22 & 0.75\\
  nc.p.c03 & No & prograde  &  0.3 & No  & 3.18 & 0.34 & --\\
  nc.p.c05 & No & prograde  &  0.5 & No  & 2.88 & 0.33 & --\\
  nc.r.c03 & No & retrograde  &  0.3 & Yes  & 1.54 & 0.28 & 0.83\\
  nc.r.c05 & No & retrograde  &  0.5 & Yes  & 1.15 & 0.26 & 0.62\\
   
 \hline
 \end{tabular}
  \caption{Properties of the models. In all models with a merger, the dwarf galaxy has the same initial mass distribution. The model nomenclature follows the properties shown in columns 2-4. The second column indicates whether the MW-like galaxy had an early clumpy phase (``c") or not (``nc"). From the third to the fifth column, we indicate the orbit configuration, prograde (``p") or retrograde (``r"), initial circularity of the merging galaxy (e.g. ``c03" or ``c05"), and whether the MW-like galaxy developed a chemical bimodality in the disk at the end of the run, at $t=10$ Gyr, respectively. Models marked with a ``*" have a  ``weak" signature of bimodality in the [O/Fe] distribution (see Section \ref{sec:chembi}). Finally, we provide a measure of the scale length, $h_{R}$ and scale heights, $h_{z1}$, $h_{z2}$, of the disk at $t=10$ Gyr. We note that models nc.r.c03, c.r.c03 are models FB80\_D1 and FB20\_D1 in \citet{amarante+2022}, respectively. }
  \label{tab:models}
  \end{center}
\end{table*}
\begin{figure}
\centering

    \includegraphics[width=\linewidth]{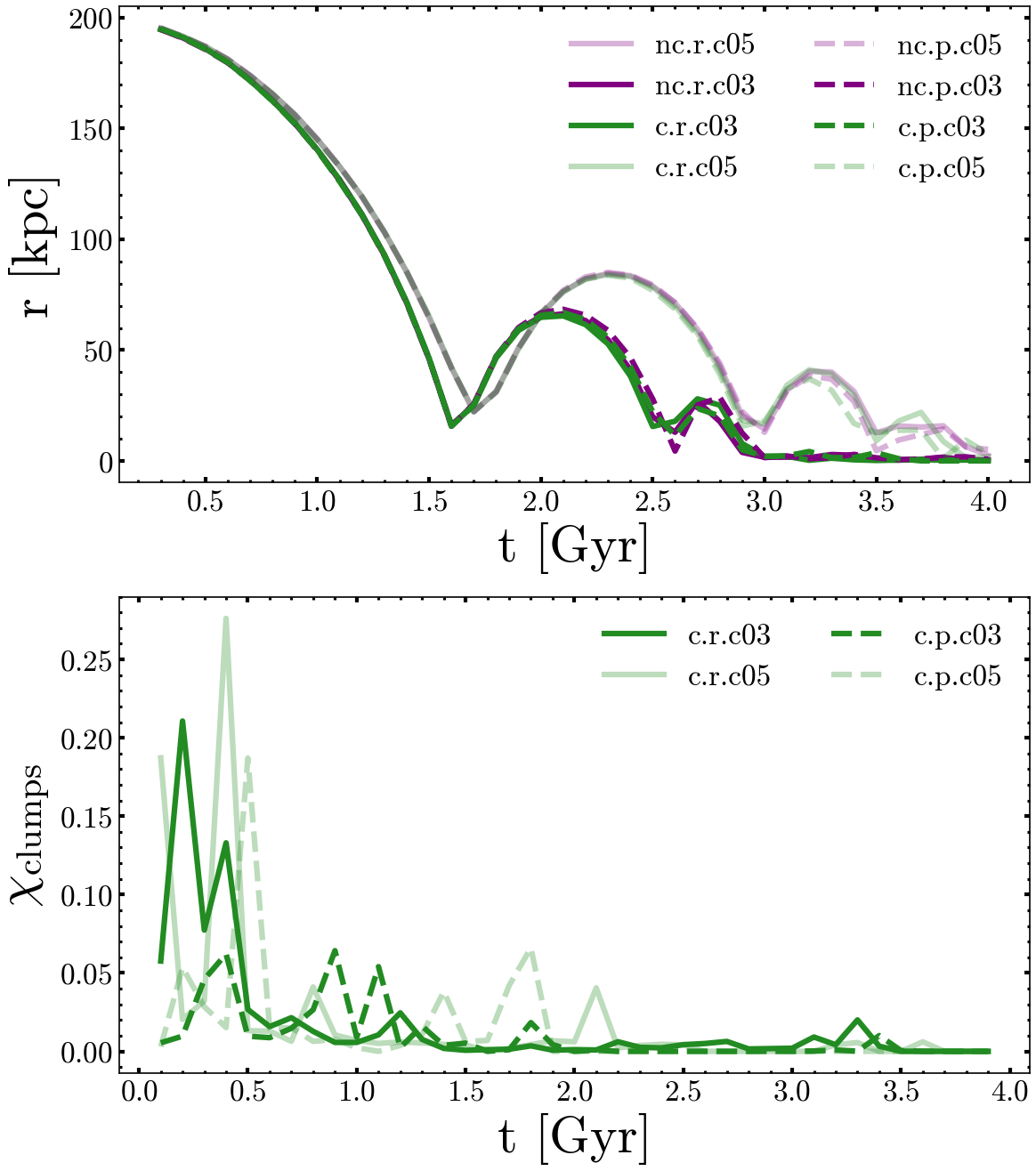}
    
    \caption{{\bf Top:} The orbital evolution of the satellite galaxy for all the merger models. Low and high supernova feedback models are shown by the green and purple lines, respectively. Prograde and retrograde mergers are represented by dashed and solid lines, respectively. While the subgrid physics and nature of the merger do not influence the orbital evolution, the initial orbital circularity does. More radial orbits (dark lines), $\eta=0.3$, have two apocenters, and the pericenters happen at earlier times compared to the merger on a more circular orbit, $\eta=0.5$.  {\bf Bottom:} The clump mass fraction evolution. The clumps, which only form in the low feedback regime, correspond to 10-20\% of the total disk mass in the first 0.5 Gyr; they cease to contribute at around $t=3$ Gyr.  }
    \label{fig:orb-clump}
\end{figure}

\begin{figure*}
\centering

    \includegraphics[width=\linewidth]{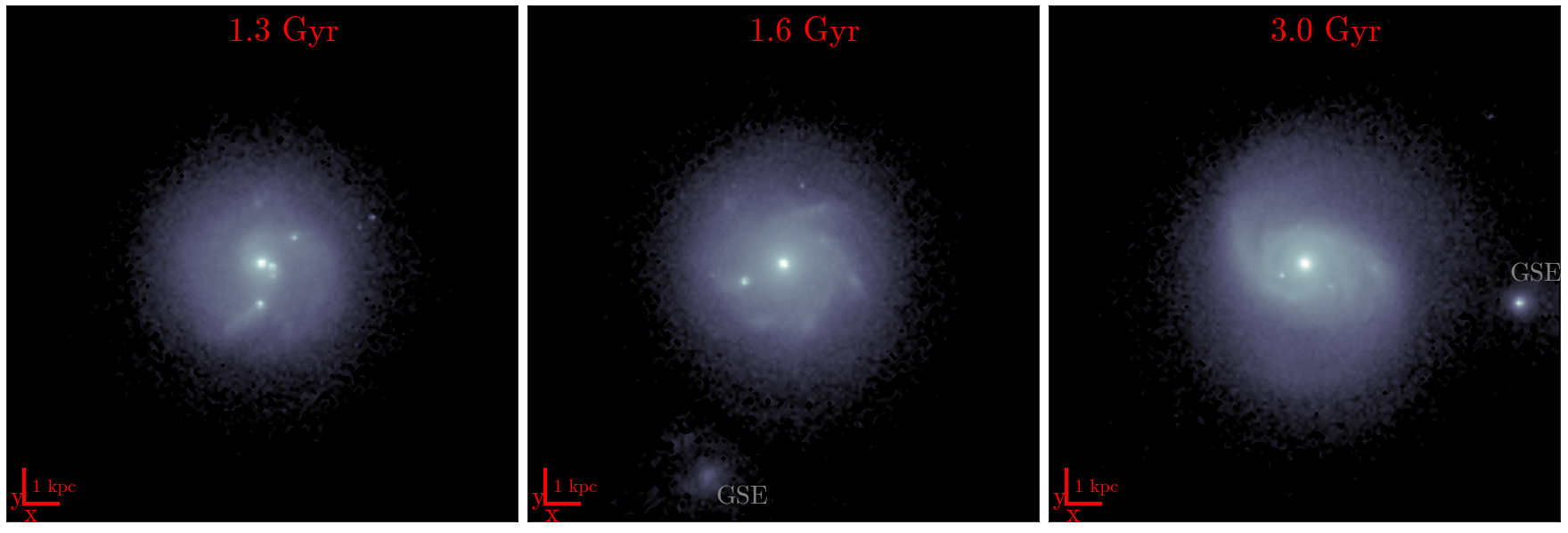}
    
    \caption{Face-on view of the stellar density distribution in model c.r.c03. Time increases from left to right and is measured since the start of the simulation at $t=0$ Gyr. The clumpy features are visible early in the evolution of the main galaxy and last up to approximately 3 Gyr. At 1.6 and 3.0 Gyr, the GSE-like dwarf is at its first and nearing its last pericenter, respectively.}
    \label{fig:visual}
\end{figure*}

We consider a total of eight models with a single merger employing the same dwarf galaxy, varying only its initial orbit and the supernova feedback of the whole model. The latter allows or prevents the formation of early clumps in the host disk (see Section \ref{sec:gridphyics}). We summarize the various key properties of these runs in Table \ref{tab:models}. As a control sample, we also consider two isolated MW-like galaxy runs with and without an early clumpy phase, ``c.iso" and ``nc.iso" respectively. All the models are evolved to $t=10$ Gyr, and we save snapshots every $t=0.1$ Gyr.  \par 
The top panel of Figure \ref{fig:orb-clump} shows the evolution of the dwarf's orbit for all the merger experiments. The time of the pericenter passage depends solely on the initial orbit circularity: dwarfs on more radial orbits, $\eta=0.3$, shown by the darker lines, have an early pericenter and two clear apocenters compared to those with $\eta = 0.5$. In other words, at these early times, the type of merger, i.e. whether prograde or retrograde, and whether the host is clumpy or non-clumpy, does not influence the orbit evolution of the merger. This will be important in the discussion presented in Section \ref{sec:discussion}\par
The bottom panel of Figure \ref{fig:orb-clump} shows the evolution of the clump mass fraction, $\chi_{{\rm clumps}}$, for models with low supernova feedback (see Section \ref{sec:gridphyics}), following the procedure described in \citet{fiteni2021} to identify the clumps and measure their mass. In this work, we calculate $\chi_{{\rm clumps}}$ relative to the total stellar mass of the disk, defined as the region within $3<{\rm R_{gc}/kpc} < 12 $ and $ |z|<3 \,{\rm kpc}$. All the models have $\chi_{\rm clumps}\approx 0.1-0.2$ in the first 0.5 Gyr of evolution and then oscillate to smaller values around $\chi_{\rm clumps}\approx 0.05$ for up to the first 2-3 Gyr. Each clump has a typical mass ranging from $ 0.3 - 1 \times 10^8\,{\rm M_{\odot}}${, compatible with the masses measured in observed clumpy galaxies \citep[e.g.][]{livermore+2012, cava+2018, huertas-company+2020, sok+2025}. Although not explicitly measured in the current work, in these tailored models, a clump typically survives for $\approx 35-300$ Myr \citep{garver+2023}. \par
Figure \ref{fig:visual} shows the stellar density at different times, illustrating distinct phases of clump formation and the dwarf galaxy orbit for model c.r.c03: before (t=1.3 Gyr), during (t=1.6 Gyr), and approaching the final (t=3.0 Gyr) pericenter passage to the MW-like galaxy.
Although the clump mass fraction decreases abruptly for $t>1$ Gyr, they are still clearly visible in the stellar density map. The GSE-like satellite can also be seen as a relatively diffuse stellar density distribution until before its first pericenter, and as a high-density stellar core as it evolves towards its last pericenter (see section 3.2 of \citealt{amarante+2022} for details on the dwarf structure evolution). \par
All of the models result in disk galaxies, qualitatively similar to the MW, but without a bar. Those that went through a retrograde merger have a smaller disk compared to those with a prograde merger (see Table \ref{tab:models}). The retrograde merger removes angular momentum from the gaseous and stellar disk, preventing the disk from growing. Nonetheless, the rotation curve of the models with a merger, shown in Figure \ref{fig:rotcurve}, is qualitatively similar to their isolated counterparts, shown as dashed lines, and comparable to the MW \citep[e.g.][]{mcmillan2017}. In the coming section, we take a closer look at how a dwarf galaxy affects the chemistry of the disk of clumpy and non-clumpy galaxies.  
\section{Creating a chemical bimodality}\label{sec:chembi}
\begin{figure*}
\centering

    \includegraphics[width=\linewidth]{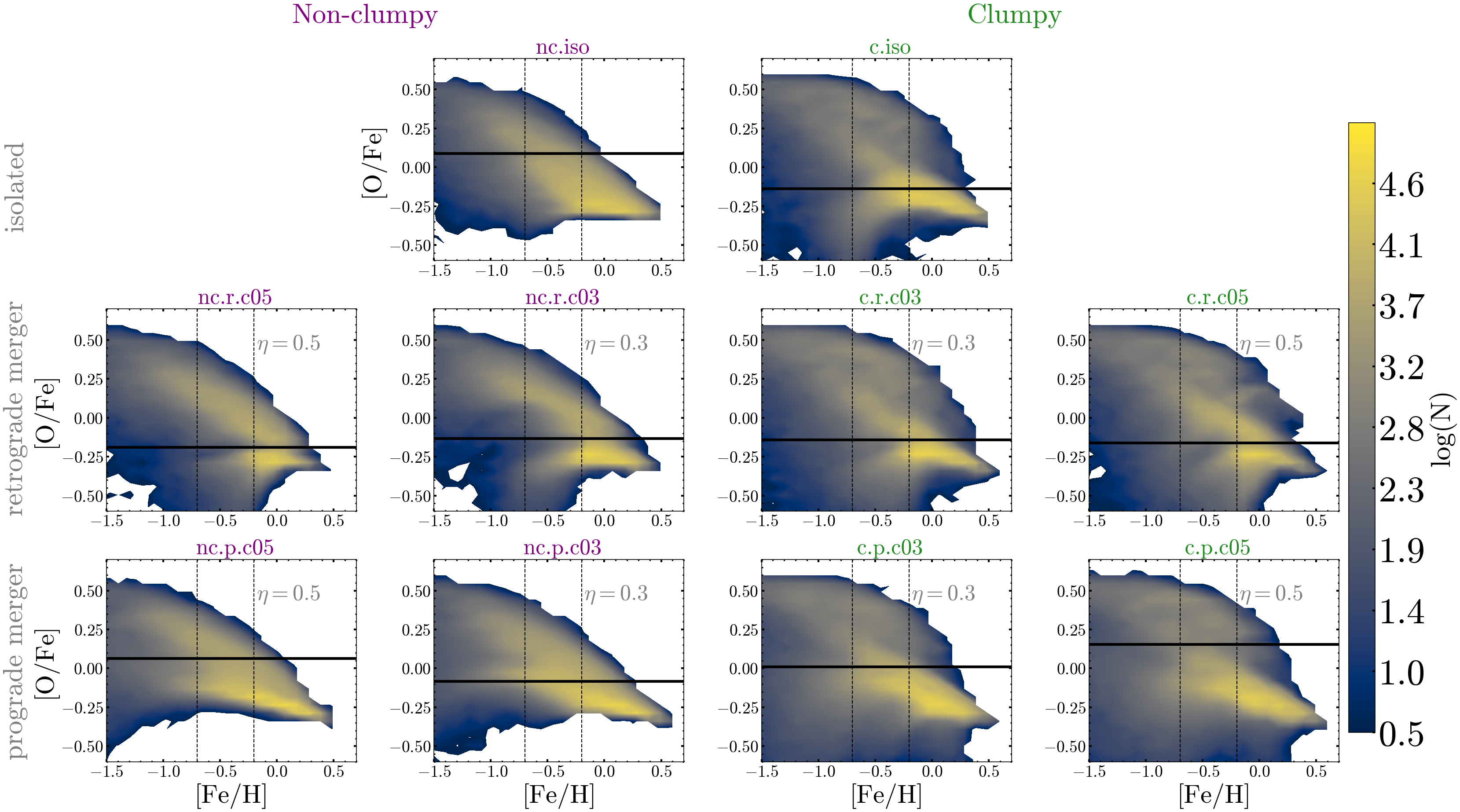}
    
    \caption{Density in the [$\alpha$/Fe]-[Fe/H] plane for the spatial range $4 < R/{\rm kpc} < 12$, and $|z|<3\, {\rm kpc}$. All models share the same initial gas and dark matter distribution for the MW-like and dwarf galaxies. Clumpy and non-clumpy models are distinguished by titles colored in green and purple, respectively. Isolated, retrograde, and prograde models are shown in the first, second, and third rows, respectively.
    The satellite's initial orbit circularity is set to 0.3 (second and third columns) or 0.5 (first and fourth columns).  All the clumpy models, including the isolated one, develop a chemical bimodality in the disk. However, the prograde mergers do not create the bimodal chemical disk in the non-clumpy model, similar to the non-clumpy isolated model. In non-clumpy models, the bimodality arises primarily from the gas-rich merger, whereas in clumpy models, it is predominantly driven by star formation in the clumps. The black dashed lines show the [Fe/H] interval on which the [O/Fe] probability density function was built (see Figure \ref{fig:ofe1d}) to separate both populations. The black solid horizontal line indicates the [O/Fe], which distinguishes the $\alpha-$populations (see Section \ref{sec:ofe}).}
    \label{fig:alphairon}
\end{figure*}

\begin{figure*}

\centering\includegraphics[width=\linewidth]{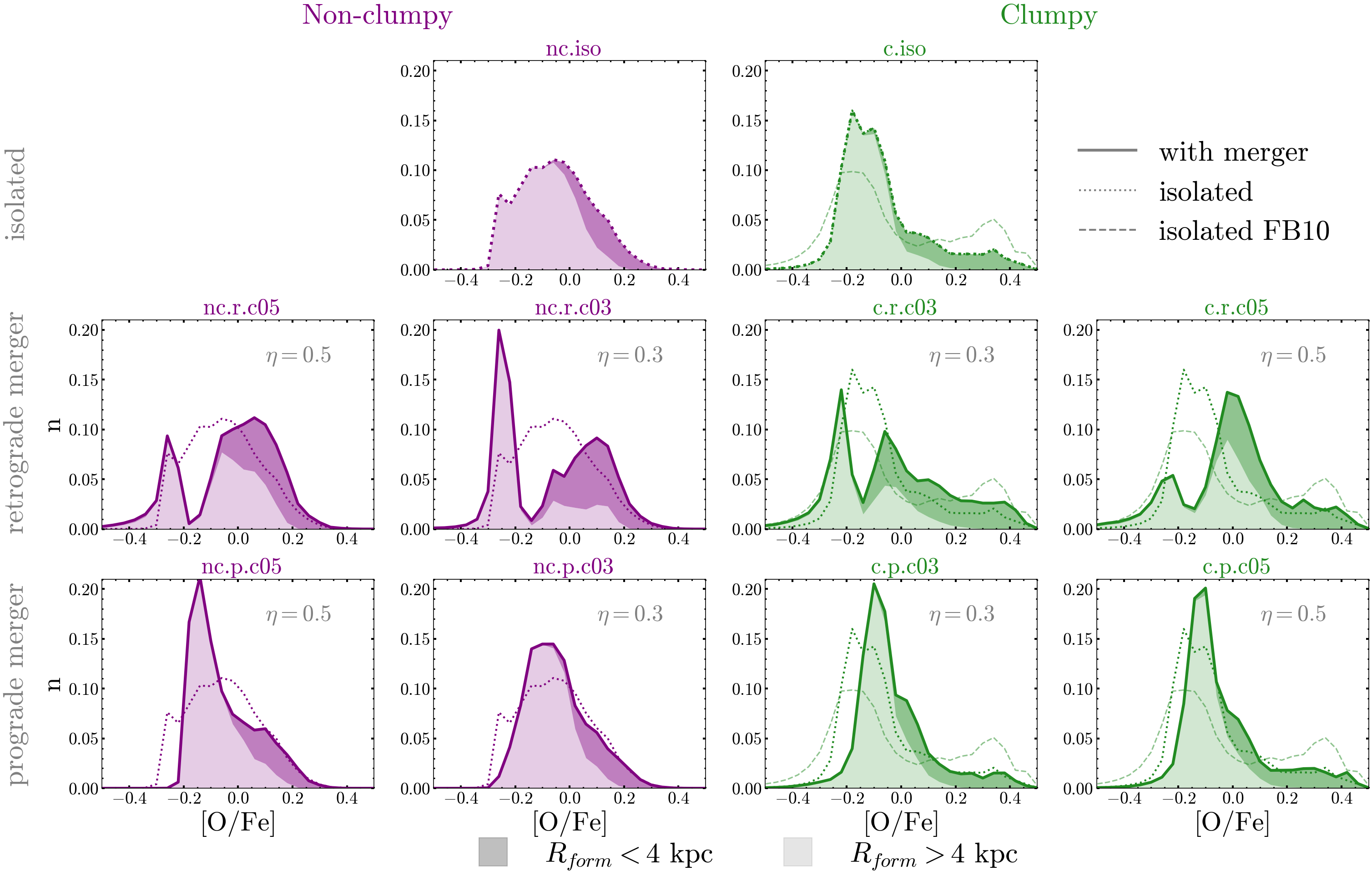}  
    \caption{[O/Fe] histogram for the interval  $-0.7< {\rm [Fe/H]} < -0.2$, $4 < R/{\rm kpc} < 12$, and $|z|<3\, {\rm kpc}$. All models share the same initial gas and dark matter distribution for the MW-like and dwarf galaxies. Clumpy and non-clumpy models are distinguished by titles colored in green and purple, respectively. Retrograde and prograde models are shown in the second and third rows, respectively.
    The satellite's initial orbital circularity is set to 0.3 (second and third columns) or 0.5 (first and fourth columns). Dotted lines correspond to the isolated control models shown in the top row. We also include an isolated model with a stronger clumpy phase (``isolated FB10") shown as the dashed line. While all the retrograde merger models can produce a double-peak distribution in [O/Fe], the non-clumpy models with a prograde merger have a similar unimodal distribution compared to their isolated counterpart. Moreover, the clumpy+prograde merger has a smaller fraction of $\alpha-$rich stars, compared to the clumpy isolated model. The initial orbital circularity of the satellite, $\eta$, is indicated in each panel. The dark (light) shaded area corresponds to the contribution of stars born in the inner (outer) parts of the galaxy, i.e. $R < 4$ kpc ($R > 4$ kpc).}
    \label{fig:ofe1d}
\end{figure*}
The bimodality in the [Fe/H]-[$\alpha$/Fe] plane is a key feature of the MW's disc \citep[e.g.][]{fuhrmann1998,venn+2004,bacham+2006,bensby+2014,anders+2014,hayden2015, queiroz+2020, imig+2023} and sensitive to its mass assembly, chemical enrichment, and stellar nucleosynthesis; thus, we take advantage of these tailored models to understand the mechanisms that may have produced it in the MW. We use oxygen as a tracer of $\alpha-$abundance and, throughout this work, we use the ratio [O/Fe] to distinguish $\alpha-$rich and $\alpha-$poor stars. Despite this ratio not being quantitatively similar to what is observed in the MW, it can still be used qualitatively to distinguish the two populations. Finally, we note that, in this work, we are interested in the shape of the chemical bimodality at the disc, encompassing the Solar Neighbourhood region and excluding the central parts of the galaxy, with two overlapping $\alpha-$sequences in metallicity. This region is defined by $4 < R/{\rm kpc} < 12$, $|z| < 3\, {\rm kpc}$.  \par 
Figure \ref{fig:alphairon} shows the chemical [Fe/H]-[O/Fe]\footnote{In all the models, the $\alpha-$abundance is tracked by Oxygen and depends strongly on imperfectly known stellar yields \citep{buck+2021}. Thus, we focus on qualitative trends in [O/Fe] rather than on precise values.} plane, {at $t=10$ Gyr}, as a 2D density distribution for the 10 models used in this work.
The isolated control models, i.e., clumpy and non-clumpy, are shown in the top row. As already shown in \citet{clarke+2019}, the isolated clumpy model forms a disk with qualitatively similar chemical bimodality as the MW, whereas the non-clumpy isolated model has a single chemical sequence.
In both cases, there is a smooth transition between the $\alpha-$enriched stars to the more $\alpha-$poor stars as the medium enriches more in iron peak elements due to a higher contribution of supernova type Ia yields over supernova type II \citep[e.g.][]{tinsley1980, nomoto+2013, kobayashi+2020}. \par
The second row of Figure \ref{fig:alphairon} shows the models that went through a retrograde merger, which is the favored orbit of the GSE satellite during its merger event \citep{koppelman2020, naidu2021}. The second and third columns show the realizations that had a more radial merger compared to the ones shown in the first and fourth columns. We observe that the chemical bimodality arises in both clumpy and non-clumpy hosts. The merger-induced formation of a geometrical thick disk from an initially thin disk has been extensively studied in N-body only models without star formation \citep[e.g.][]{quinn+1993,villalobos-helmi2008}. However, our attention here is focused on how a merger induces the formation of the disk's chemical bimodality with SPH+N-body models.   \par
The third row of Figure \ref{fig:alphairon} shows the chemical plane of the disk stars for the models that had a prograde merger. We see that both non-clumpy and clumpy realizations can develop a double sequence, although we show in the next section that it is not as pronounced as in the models with a retrograde merger. In fact, recently \citet{johnson+2025} demonstrated with analytical galactic evolution models that a prograde merger fails to induce a chemical bimodality in the disk.
\par
\subsection{The [O/Fe] distribution}\label{sec:ofe}
We now investigate the distribution of [O/Fe] over the annulus defined by $4 < R/{\rm kpc} < 12$, $|z| < 3\, {\rm kpc}$, and $-0.7 < {\rm [Fe/H] < -0.2}$. This spatial range encompasses the geometrical thin and thick disks, which are associated with the $\alpha-$poor and $\alpha-$rich populations \citep[e.g.][]{hayden2015, imig+2023}.\par
Figure \ref{fig:ofe1d} shows the chemical bimodality in the [O/Fe] distribution for disk stars within the aforementioned spatial and chemical range. The clumpy and non-clumpy models are shown in green and purple, respectively. For comparison, the distributions of the control models, i.e., without a merger (shown in the top row), are overplotted as a dotted line alongside their counterparts that include a merger (i.e., whether it is clumpy or non-clumpy).  Although all models with a retrograde merger (second row) have a clear [O/Fe] bimodality, the relative heights of the peaks appear to depend on both the merger’s orbital circularity and whether the host galaxy underwent an early clumpy phase. \par
The bottom row shows the [O/Fe] for models that had a prograde merger. Differently from the retrograde scenario, a prograde merger does not create a clear [O/Fe] bimodality, i.e. a distribution with two distinct peaks and a well-defined trough.
Instead, a prograde merger tends to favor the formation of a more prominent low-$\alpha$ sequence with an extended [O/Fe]-rich tail, so that the [O/Fe] distribution resembles that of the isolated clumpy model (c.iso). This effect is enhanced in the models with an early clumpy star formation phase. We also compare the clumpy+merger models with the isolated clumpy model of \citet{clarke+2019} (green dashed line), which had reduced supernova feedback efficiency in its coupling with the gas\footnote{10\% of the $10^{51}$ erg injected per supernova.}. This model produced a more intense clumpy phase, reinforcing the observation that a stronger clumpy phase enhances the contribution of the $\alpha-$rich disk. \par
Since all the models start with the same structural initial conditions for the dwarf -- varying only in orbital parameters -- the total gas mass brought by the merger is approximately similar across all the models. In fact, we verified that only $\approx 5-10\%$ of the dwarf's original gas settles in the MW's disk after the second pericentric passage. However, differences in gas deposition patterns and the amount of dynamical heating may either promote or suppress chemical bimodalities. 

Finally, we define how we separate the $\alpha-$rich and $\alpha-$poor populations in each model. Using the [O/Fe] distribution within the previously defined spatial and [Fe/H] ranges, we build a Gaussian Kernel Density Estimate (KDE) with bandwidth $=0.05$ dex. We then evaluate the probability density function from the KDE and identify the local minimum in the density curve. This minimum defines the separation between the $\alpha-$rich and $\alpha-$poor populations (solid black line in Figure \ref{fig:alphairon}). This definition is adopted for all analyses presented in the remainder of the paper. In the following sections, we investigate the specific mechanisms driving the chemical bimodalities in the models. \par
\subsection{The star formation rate}\label{sec:sfr}

\begin{figure*}
\centering

    \includegraphics[width=\linewidth]{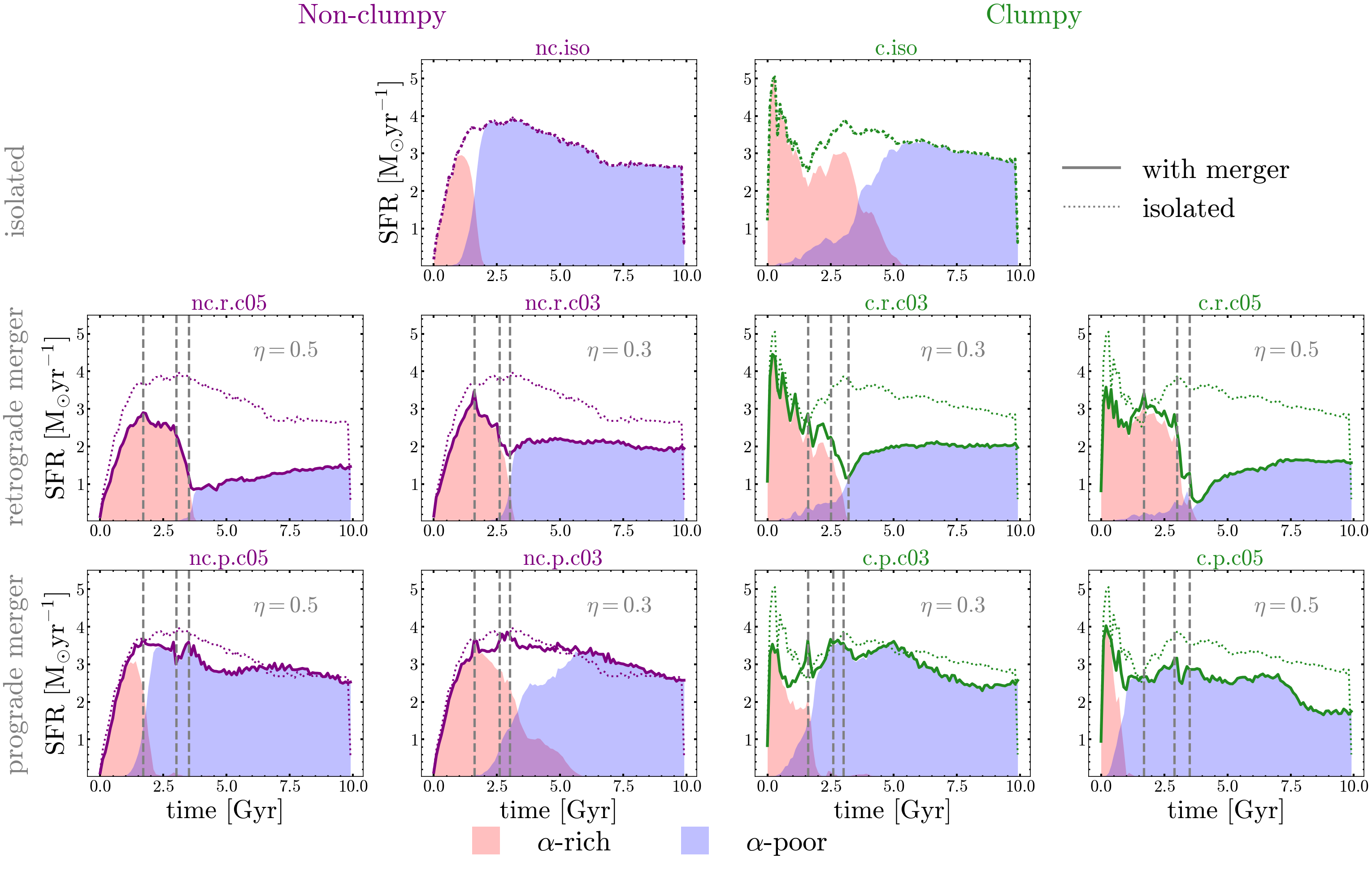}
    \caption{The SFR of the disk, $4 < R/{\rm kpc} < 12$, and $|z|<3\, {\rm kpc}$. The solid line shows the case for the isolated models (top row), and for those with a retrograde (middle row) or prograde merger (bottom row). Purple and green titles correspond to non-clumpy and clumpy models. The contribution to the SFR from the $\alpha-$rich and $\alpha-$poor populations is represented by orange and blue areas, respectively. The models with a clear chemical bimodality have two main features: {\it i)} a significant decrease in SFR and {\textit ii)} the SFR of the $\alpha-$rich stars at earlier times is higher than the $\alpha-$poor population at later times. These occur in the isolated clumpy models and the non-clumpy and clumpy models with a retrograde merger. A prograde merger, however, does not significantly reduce the $\alpha-$poor SFR, and fails to produce the chemical bimodality of the clumpy model. The vertical dashed lines indicate the time of the dwarf's pericenter passages during the merger. The initial orbital circularity of the satellite, $\eta$, is indicated in each panel. The green and purple dotted lines show the SFR for the isolated clumpy and non-clumpy models, respectively.}
    
    \label{fig:sfr}
\end{figure*}

Figure \ref{fig:sfr} shows the observed star formation rate (SFR), measured from the stellar age distribution at $t=10$ Gyr, for all the 10 models, within the spatial range $4 < R/{\rm kpc} < 12$, and $|z|<3\, {\rm kpc}$. This mimics observational measurements in the MW, but we note that it is not the actual SFR within this volume, as stellar migration can affect this measurement \citep[e.g.][]{roskar+2008,frankel+2018,minchev+2025}. The contribution from the $\alpha-$rich and $\alpha-$poor populations to the total SFR is shown as the orange and blue shaded areas, respectively. We start by highlighting three general properties regarding the SFR of our models. \par
All models exhibit a quenching of the $\alpha-$rich population at $t\approx2-3$ Gyr, independent of the orbital properties of the dwarf merger, which is followed by an increase in the SFR of the $\alpha-$poor population. Models with a merger have a sudden but small increase in the SFR during the pericentric passages of the satellite, indicated by the vertical dashed lines. Instead, the $\alpha-$rich population has a bursty SFR regulated by the clumps in all the clumpy models. \par
The top row shows the SFR for the isolated models in the clumpy and non-clumpy scenarios. Besides the aforementioned decline in SFR, these two isolated models also differ in the relative difference in the SFR between the $\alpha-$poor and $\alpha-$rich stellar populations. In the clumpy scenario, which exhibits chemical bimodality, the disk's $\alpha-$rich population has a higher peak in the SFR compared to the peak of the $\alpha-$poor. In contrast, in the non-clumpy model, the SFR of the $\alpha-$rich stars peaks at a lower value compared to the $\alpha-$poor. The star formation rate thus acts as a discriminator for the origin of the chemical bimodality in simulated disks \citep{clarke+2019, beane25, orkney+2025}. \par
Here, with our tailored merger models, we can directly assess the effect of a single merger event on the disk's SFR and investigate whether a merger like the GSE can alter the SFR to the extent of creating or suppressing the formation of a chemical bimodality in the disk. The second row of Figure \ref{fig:sfr} shows the SFR for the retrograde merger models for which the $\alpha-$rich population has systematically higher SFR compared to the $\alpha-$poor, regardless of whether the host galaxy had a clumpy or non-clumpy episode. In terms of absolute values, when compared to the isolated control models, shown by the dotted lines, the retrograde merger does not significantly change the $\alpha-$rich SFR. But the $\alpha$-poor population has its SFR substantially reduced. This change in the SFR evolution produces a clear chemical bimodality in the non-clumpy and clumpy disks with a retrograde merger.  \par
The last row of Figure \ref{fig:sfr} shows the SFR for the prograde mergers. In the non-clumpy models (first two columns), the merger does not significantly reduce the SFR for the $\alpha-$poor population, as in the retrograde merger. This prevents the emergence of the disk's chemical bimodality. On the other hand, the clumpy models reveal a more complex behavior: the prograde merger reduces the SFR of the $\alpha-$rich population, compared to the isolated clumpy model. This leads to both a reduced fraction of $\alpha-$rich stars in the disk and a weaker chemical bimodality, compared to the retrograde merger scenario. \par%
Our analysis of the star formation rates in these tailored models reveals that a single merger event can both establish or suppress the formation of the chemical bimodality signatures in a galaxy's disk at the solar neighborhood. We further explore the birth location of the stars contributing to the chemical bimodality in the disk in the next section. 
\subsection{The birth radii from the stars creating the chemical bimodality}\label{sec:migration}
We start this subsection by clarifying an important concept used here. We refer to ``migration" as any process that causes a star to end up located away from its formation radius, $R_{form}$, at the present day. We stress that this includes both the process of churning, which is the change of the stars' $L_z$ without heating their orbits radially, and blurring, which increases the orbit eccentricity \citep{sellwood-binney2002}, and the rapid orbit scattering caused by clumps.  We defer to future work quantifying the amount of churning and blurring in the chemical disks caused by a single merger. Nonetheless, we note that \citet{beraldo+2021} found that stars in near-circular orbits in the $\alpha-$rich and $\alpha-$poor disk go through comparable churning in an isolated MW-like galaxy. \par
We return to Figure \ref{sec:chembi}, where we measured the present day [O/Fe] distribution for stars at $4 < R_{form}/{\rm kpc} < 12$ and $|z| < 3$ kpc, and distinguish the contribution of stars that were born in the inner parts of the galaxy, $R_{form} < 4$ kpc, from those born at $R_{form}>4$ kpc, as dark and light shaded areas, respectively. In the non-clumpy isolated model, a significant fraction of $\alpha-$rich stars were born in the inner galaxy and migrated to our volume of interest, but these migrated stars alone are not able to create the chemical bimodality in the disk. In the isolated clumpy model, the migrated stars create an extended tail towards high [O/Fe], but a bimodality forms only when the clumpy phase is intense enough, as shown by the dashed line.  \par
The models with a retrograde merger, which exhibit the most pronounced chemical bimodality, demonstrate that radial migration can amplify an existing bimodality by transporting $\alpha$-rich stars from the inner galaxy, $R < 4$ kpc, to the outer regions. Stars born at $R_{form}>4$ kpc already form a double-peaked [O/Fe] distribution with a discernible valley, as indicated by the light-shaded area. Furthermore, we find that among the migrated $\alpha-$rich stars, approximately 50\% are on circular orbits, $\eta>0.9$, as shown in Figure \ref{fig:ofe1dcirc}. \par
On the other hand, models with a prograde merger show that migration alone is not enough to create a clear chemical bimodality. For instance, stars born locally (i.e. at $R_{form}>4$ kpc) do not show a double-peaked [O/Fe] distribution, and the migrated stars are not enough to create a clear $\alpha-$rich peak, although they can enhance it. This contrasts with conclusions drawn from pure analytical models, where migration has been proposed to directly create a chemical bimodality \citep[e.g.][]{schonrich-binney2009_chemdisc, sharma+2021}.\par
Finally, we note that the amount of stars migrating outwards in the MW should be affected by the presence of the bar \citep[e.g.][]{minchev-famaey2010,minchev+2012,chiba-schonrich2021, marques+2025}, which our models do not have. Nonetheless, it remains to be seen if migration caused solely by the bar would be enough to create a chemical bimodality in the disc. \par

\section{The transition to an $\alpha-$poor disk}\label{sec:age}
We showed, in the previous section, how a single merger and massive star-forming clumps can modulate the SFR  of both the $\alpha-$rich and $\alpha-$poor populations, affecting the development of any chemical bimodality. In this section, we study how this modulation can imprint distinct signatures in the age distribution for both populations and set the time at which the fraction of $\alpha-$poor stars is higher than the $\alpha-$rich stars in the disk. We refer to this time as the transition time. \par
A key imprint of the clumpy models is the presence of a low SFR for the $\alpha-$poor population as early as $t\approx 1$ Gyr, which promotes the formation of the old (age $>10$ Gyr) $\alpha-$poor population. This contrasts with the non-clumpy models where the $\alpha-$poor SFR starts at $t>2$ Gyr, and can be delayed up to $t\approx 3$ Gyr in the retrograde merger scenario (Figure \ref{fig:sfr}, second row, first and second columns). The early formation of the $\alpha-$poor disc has been discussed in \citet{beraldo+2021} with isolated models, and we will further explore it in the next section. \par
Figure \ref{fig:fraction} shows the fraction of $\alpha-$poor and $\alpha-$rich stars at a given age bin (similar to figure 4 in \citealt{borbolato+2025} for the MW). The isolated models are shown as the dotted lines. The clumpy isolated model has a significant fraction, $>10\%$, of $\alpha-$poor stars as old as 11.5 Gyr, which is a direct result of its SFR discussed in Section \ref{sec:sfr}. In fact, model iso.fb10 (colored dashed lines), which has a stronger clumpy phase, has a fraction of low$-\alpha$ disk stars older than $11$ Gyr as high as $20\%$. \par
As can be seen in Figure \ref{fig:fraction}, the time of transition depends in a non-trivial way on the orbit of the merger and the clumpy phase of the main galaxy. Nevertheless, some important qualitative insights are apparent. Firstly, a retrograde merger in a non-clumpy galaxy, besides creating a chemical bimodality (see Section \ref{sec:chembi}), also delays the transition to an $\alpha-$poor dominated disk, i.e. when its fraction is greater than $50\%$, which occurs $\approx 0.5$ Gyr after the last pericenter of the dwarf. On the other hand, in the clumpy models with a retrograde merger, the transition happens earlier compared to the clumpy isolated models, although coincidentally, still at a similar time. \par
The delay and anticipation of the transition in non-clumpy and clumpy models, respectively, also occur in the prograde merger models when the circularity is low, i.e. $\eta=0.3$. However, in the non-clumpy model, the merger does not affect the transition time to an $\alpha-$poor disk, while for the clumpy model, a prograde merger on a more circular orbit advances the transition compared to its isolated model counterpart. Nonetheless, the transition time is similar to the isolated model with a strong clumpy phase.\par
These results suggest that the transition to an $\alpha-$poor dominated disk can be regulated by a single merger event depending on the satellite's orbit. This is similar to what is observed in cosmological simulations, where mergers can tidally induce pristine/metal-poor gas infall, resulting in the formation of a low$-\alpha$ disk \citep{renaud+2021,orkney+2025}. We note that this is also degenerate with how clumpy the star formation is in the host galaxy.
\begin{figure*}
\centering

    \includegraphics[width=\linewidth]{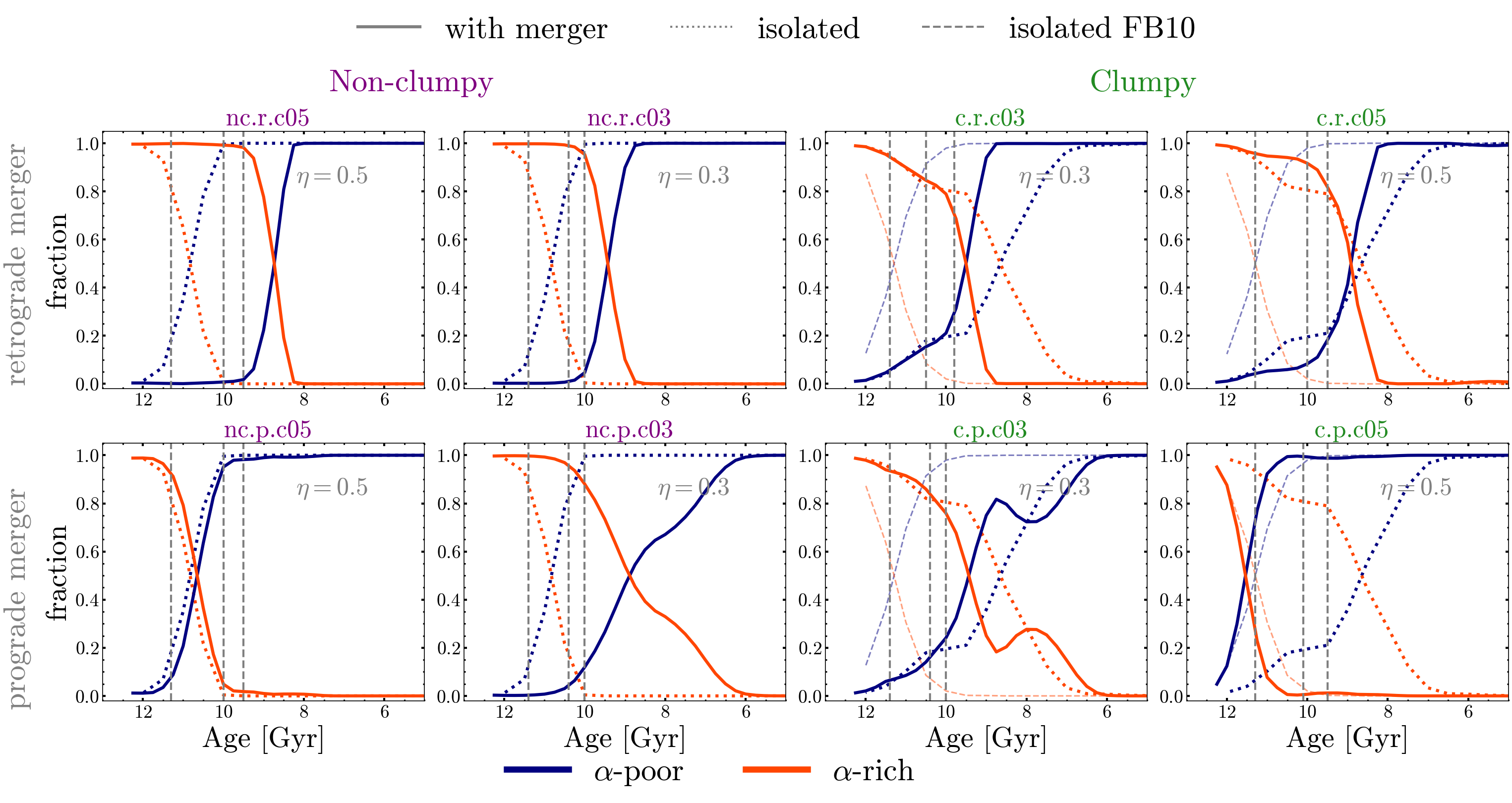}
    
    \caption{Present-day chemical disks mass fraction per age bin. We use the previous definition of $\alpha-$poor (blue) and $\alpha-$rich (orange) populations to calculate their fraction contribution in each age bin. Colored dotted and dashed lines show the corresponding clumpy or non-clumpy isolated model counterpart. The time of the pericenter passage is indicated by the vertical gray dashed lines.  }
    \label{fig:fraction}
\end{figure*}
\section{The old $\alpha-$poor disk}\label{sec:old}
The properties of the $\alpha-$poor and $\alpha-$rich disks have been extensively explored in the literature (see e.g. \citealt{imig+2023} for a recent overview). For instance, their chemical abundances \citep[e.g.][]{bensby+2003, prantzos+2023}, kinematics \citep[e.g.][]{yan+2019,hu+2022, hu+2023}, ages \citep[e.g.][]{wu+2023,fernandez-alvar+2025, cerqui+2025}, and spatial distribution \citep[e.g.][]{bovy+2012,lian+2022} highlight distinct formation processes. The current view of the formation timeline of the MW disk can be summarized as follows:
\begin{itemize}
    \item[i.] An early turbulent phase of the MW's history, during which metal-poor stars dominated the proto-Galaxy in a pressure-supported system \citep{belokurov+2022, conroy+2022, rix+2022};
    \item[ii.] This is followed by the rise of angular momentum as the Galaxy enriches in [Fe/H], a process termed the spin-up phase, turning it into a rotationally supported system \citep{belokurov+2022, conroy+2022, xiang-rix2022,chandra+2024, viswanathan+2024};
    \item[iii.] The merger with the GSE dwarf, which quenched the $\alpha-$rich disk and sets the start of the SFR for the $\alpha-$poor disk \citep{gallart+2019, bonaca+2020}. 
\end{itemize} 
This sequential scheme also aligns qualitatively with several MW analogues in cosmological simulations \citep[e.g.][]{buck+2020}. However, it fails to explain the presence of a significant fraction of old $\alpha-$poor stars on circular orbits \citep{silva-aguirre+2018, laporte+2020, beraldo+2021, nepal+2024, gent+2024, borbolato+2025}. Recently, \citet{borbolato+2025} analyzed $\approx 60$ thousand stars with spectroscopic information from LAMOST and APOGEE data and reliable age measurements from {\tt StarHorse} \citep{basilio+2016, queiroz+2018}. They found a relatively large number of old (age $>$ 11 Gyr) $\alpha-$poor disk stars. This reinforces that this population, although a small fraction of the entire $\alpha-$poor disk, represents a non-negligible fraction of old disk stars. Thus, its presence needs to be better understood theoretically and reproducible through Galactic chemical evolution and galaxy formation models.\par
\cite{beraldo+2021} concluded that the clumpy scenario is a plausible formation mechanism for the old $\alpha-$poor population. They also showed that an isolated MW-like galaxy with an early clumpy phase could reproduce the peaks of the observed pericenter distribution of RR Lyrae stars, which are typically old \citep[but see e.g.][on the possibility of relatively young RR Lyrae stars]{iorio-belokurov2021,bobrick+2024, zhang+2025, mateu+2025}, on disk orbits in the MW \citep{prudil2020}. Now, with the models presented here, we can further explore whether a GSE-like merger can impact the formation of the old $\alpha-$poor disk.  \par
Similar to their isolated counterparts, a clumpy galaxy which underwent an early merger also has a non-negligible number of old $\alpha-$poor stars. From Figure \ref{fig:fraction} we can see that there are between 5-20\% $\alpha-$poor stars in the oldest age bins, i.e. age$>$10 Gyr\footnote{As our simulations were run up to $t=10$ Gyr, we added +3 to the ages of the stars in our models. We note that this is the same as selecting stars with formation time $t_{form} < 3$ Gyr. This age shift allows a better comparison with observational data}. \par
Interestingly, the merger affects the presence of old $\alpha-$poor stars in the solar cylinder for the non-clumpy models. The isolated non-clumpy model and its counterparts with the prograde mergers on a more circular orbit have a significant fraction of old $\alpha-$poor stars. However, these models do not have a chemical bimodality (see Fig. \ref{fig:ofe1d}); thus, they are not strictly similar to the MW. The non-clumpy models with a retrograde merger, which have a chemical bimodality, do not have an old $\alpha-$poor population at the solar cylinder. In fact, a merger in a non-clumpy model can either delay the transition from high$-$ to low$-\alpha$ disk, or conversely, completely prevent the formation of an old low-$\alpha$ disk by transitioning very early. \par
We can conclude that, in these tailored models, although a merger can induce the formation of chemical bimodality, an early clumpy formation is necessary to form a significant fraction of old $\alpha-$poor stars at the solar cylinder at the same time that it produces a chemical bimodality. In fact, a stronger clumpy phase produces a relatively larger fraction of old $\alpha-$poor stars in the oldest age intervals, i.e. age $>11$ Gyr, as can be seen by the blue dashed lines in the last two columns of Figure \ref{fig:fraction}. 

\section{Discussion}\label{sec:discussion}
In this section, we discuss our results in the context of previous work using tailored and fully cosmological models (Sections \ref{sec:beane} and \ref{sec:cosmo}, respectively). Finally, we explore the connection between our models with clumpy galaxies observed at redshift $z\approx 2$ in Section \ref{sec:linkz}. Throughout, we also discuss the caveats of our experiments.
\subsection{Comparison with Beane et al. (2025)}\label{sec:beane}
Tailored merger simulations that include star formation, such as those presented here, enable us to explore a larger parameter space, such as the orbit configuration of a merger and different subgrid physics, and their effect on a MW-like galaxy. This comes at a relatively lower computational cost compared to full cosmological simulations. In the context of the impact of a GSE-like merger on the chemistry of the MW disk, \citet{beane25} presented a suite of tailored N-body models, which included gas and star formation, of MW-like galaxies that went through a GSE-like merger during their first Gyr of evolution. They propose that a brief, $t< 300$ Myr, halt in star formation can lead to a bimodal [$\alpha$/Fe] distribution at a given metallicity interval. Previously, pure analytical models have also suggested a brief halt in the SFR in the MW for a chemical bimodality to emerge \citep[e.g.][]{chiappini+1997}. In \citet{beane25}, the GSE merger event can trigger this halt, but they noted that non-merger scenarios can also present an SFR quenching. \par
Contrary to \citet{beane25}, our models with a chemical bimodality in the disk do not completely quench. This is in line with current observational measurements showing that the MW disk had a continuous star formation history, with a concomitant period of formation of the $\alpha-$poor and $\alpha-$rich sequences \citep{ratclife+2025}. As discussed in Section \ref{sec:sfr}, our models have a significant decrease in the total SFR in the disk either due to the end of the clumpy phase (model c.iso) or induced by the retrograde merger in both clumpy and non-clumpy models, but the disk is always forming stars. 
 \par
Another difference is the time of the first pericenter passage of the satellite: \citet{beane25} tune it to happen at an earlier time, $t\approx 0.8$ Gyr, compared to the ones presented here, $t\approx 1.6$ Gyr. Nonetheless, we argue that the decrease, or in their case the halt, in the SFR happens at the time of complete disruption of the satellite, in both suites of models. \par
One last difference is the presence of a significant fraction of old (age $>10$ Gyr) $\alpha-$poor population found in our clumpy models (Section \ref{sec:old}). While some of the models presented in \citet{beane25} form the chemical bimodality, the SFR of the $\alpha-$poor population only starts after the $\alpha-$rich population ceases its formation. Therefore, they are unable to reproduce the co-formation period between these populations.

\subsection{Comparison with cosmological simulations}\label{sec:cosmo}
Our experiments model a MW-like galaxy with a fixed DM halo mass of $10^{12}\, {\rm M_{\odot}}$, making them idealized relative to the $\Lambda$CDM hierarchical galaxy assembly framework. Although DM halos are expected to grow over time, their masses are defined relative to the critical density of the universe, which decreases towards lower redshift. This effect leads to a pseudo-growth of the DM halo and dominates their apparent evolution from $z=1$ to $z=0$ for halos of $10^{12}\, {\rm M_{\odot}}$ \citep[e.g.][]{diemand+2007,diemer+2013, zemp2014}. Therefore, a MW-sized galaxy will likely not experience substantial physical growth in its inner regions over this period.
Furthermore, \citet{semenov+2024} used Illustris TNG50 \citep{springel+2018, nelson+2019} to show that MW-like galaxies that went through a rapid increase (timescale $\lesssim 1$ Gyr) of their rotational support, as in the MW \citep[e.g.][]{xiang-rix2022,belokurov-kravtsov2024}, typically accreted their total mass in the first 2 Gyr of evolution. Therefore, in terms of total DM halo growth, the models presented here capture key aspects of MW-like evolution within a fully cosmological context.\par
%
Although our experiments are idealized, it is encouraging to see that we can obtain similar conclusions on the formation of the chemical bimodality in non-clumpy star formation models as in some cosmological realizations through the mediation of a merger event \citep[e.g.][]{buck+2020,renaud+2021, orkney+2025}. However, we note that having a GSE-like merger may not be a necessary requirement to form a chemical bimodality \citep[][]{orkney+2025}.\par
A common feature across several current cosmological simulations of MW-like galaxies forming chemical bimodalities \citep[e.g.][]{grand+2018, buck+2020,  parul2025, orkney+2025} is that the high$-\alpha$ and low$-\alpha$ tracks form sequentially. Alternatively, cosmological models showing a phase of co-formation do not produce a significant amount of stars older than $>10$ Gyr in the $\alpha-$poor sequence \citep{agertz2021, renaud+2021}. Observations of the MW stars, on the other hand, seem to suggest some level of coevalness between the low$-$ and high$-\alpha$ sequences during the first $\sim 3$ Gyr of evolution \citep[e.g.][]{silva-aguirre+2018, prudil2020,beraldo+2021,gent+2024,dorazi+2024, borbolato+2025} which is reproduced in our idealised models with a clumpy star formation mode. Currently, the effect of clumpy star formation in the chemodynamical evolution of MW-like galaxies has not been considered in a full cosmological context. 

\subsection{Linking the MW with clumpy galaxies at $z\approx2$}\label{sec:linkz}
Clumpy galaxies at redshift $z\approx2$ were first observed at rest-frame UV/optical with the Hubble Space Telescope ({\it HST}) \citep[e.g.][]{cowie+1995, vandenbergh+1996, elmegreen+2005}. More recently, with the {\it JWST}, it has been possible to confirm that a large fraction of these clumps, up to $\approx85\%$, observed in the rest-frame near-IR at $z>1$ overlap spatially with their UV counterpart \citep{kalita+2024, kalita+2025}.
Several key properties of the clumps have been identified with {\it HST} and {\it JWST}: their sizes and mass ranging from $100-500$ pc and $\approx 10^7-10^{8.5}$, respectively \citep[e.g.][]{dessauges-zavadsky+2017,cava+2018, ambachew+2022, kalita+2025size}, their lifetime have been estimated at a few hundred Myr \citep[e.g.][]{claeyssens+2023,sok+2025}, and they significantly contribute to the total star formation of a galaxy \citep[e.g.][]{wuyts+2012,guo2018}. Finally, due to the relatively high fraction, $\approx 30-55 \%$, of clumpy galaxies across a range of redshifts \citep[e.g.][]{guo+2015,sattari+2023,delaVega+2025}, it is reasonable to ask what the effects of clumps are in the formation of a MW-like galaxy. \par
Clumps were proposed as a possible formation mechanism of galactic bulges \citep[e.g.][]{noguchi1999,dekel+2009} and geometrical thick disks \citep{bournaud2009}. A series of works directly linked the clumpy formation in an isolated disk galaxy to several present-day observational properties of the MW, such as the emergence of the chemical bimodality of the disk \citep[and this work]{clarke+2019}, the chemo-geometrical dichotomy of the disk \citep{beraldo+2020}, the presence of a low-angular momentum, $v_{\phi}<100\,{\rm km/s}$, metal-rich, $-0.7<{\rm [Fe/H]} <-0.2$, in-situ halo population \citep{amarante+2020splash}, the presence of an old, age$<10$ Gyr, $\alpha-$poor disk \citep[and this work]{beraldo+2021}, and the chemistry of the bulge \citep{debattista+2023}. Moreover, the clumps in these tailored experiments have sizes, masses, and lifetimes similar to what is observed in high redshift galaxies \citep{garver+2023}.\par
Most of the aforementioned properties of the MW have several other formation mechanisms other than clumps, including mergers, with the exception of the presence of the old $\alpha-$poor disk population. Therefore, this work is a step forward in complexity by adding a single merger event in the clumpy scenario. As discussed throughout the text, although the merger can induce the formation of the chemical thick disk, only models with clumps have a significant fraction of old $\alpha-$poor disk stars. Therefore, in light of the presence of the early spin-up phase of the MW \citep[e.g.][]{belokurov+2022,conroy+2022, rix+2022} and the presence of the old $\alpha-$poor population, we argue that the significant fraction of MW-mass clumpy galaxies at redshift $z \gtrsim 2$ \citep{ferreira+2022, robertson+2023, ferreira+2023} are potential progenitor analogues of our Galaxy.
%
\section{Conclusion}\label{sec:conclusion}

In this contribution, we have studied the formation of the chemical bimodality in the [$\alpha$/Fe]-[Fe/H] plane of several MW-like hosts in the presence of a GSE-like merger. We have explored different orbital setups of the merging satellite and the mode of star formation that regulates the formation or absence of massive star-forming clumps in the disk during the first few Gyr of evolution. Our main conclusions are the following:

\begin{itemize}
    \item A chemical bimodality in the disk, $4 < R/{\rm kpc}<12$, naturally occurs when there is a significant decrease in the SFR for a brief period of time, typically $\lesssim 300$ Myr (Section \ref{sec:sfr}, Figure \ref{fig:sfr}). In our experiments, this is caused by:
    \begin{itemize}
        \item A retrograde merger, as it significantly decreases the SFR of the $\alpha$-rich population, allowing type Ia SNe to enrich the interstellar medium enough for the low$-\alpha$ sequence to emerge;
        \item Or, massive star-forming clumps, which have a high star formation rate in the first $\approx 3$ Gyr of evolution, creating a significant fraction of $\alpha-$rich stars. Once the clumps cease to form, there is an abrupt decline in the SFR, allowing the formation of the bimodality.
    \end{itemize}
    \item In the disks with the chemical bimodality, the SFR of the $\alpha-$rich population is higher in the first Gyr of evolution compared to the SFR of the $\alpha-$poor population at later times, in agreement with \citet{clarke+2019} based on an isolated clumpy model;
    \item A prograde merger does not seem to significantly decrease the SFR in the disk. This creates a [O/Fe] distribution with a prominent $\alpha-$poor peak and an extended [O/Fe]-rich tail which can be enhanced by an early clumpy phase (bottom panels of Figure \ref{sec:chembi}). This result, using full hydrodynamic models with star formation, agrees to some extent with predictions from chemical analytical models \citep{johnson+2025};
    \item A GSE-like merger can delay the transition from an $\alpha-$rich to an $\alpha-$poor dominated disk (Figure \ref{fig:fraction});
    \item In our models, the chemical bimodality in the disk is created from a combination of stars born in the outer, $R_{form}>4$ kpc, and inner, $R_{form}<4$ kpc, regions (Section \ref{sec:migration}). Contrary to expectations from pure analytical models, stars originating from the inner parts fail to produce the disk's chemical bimodality, but can enhance it \citep[see also][]{barry+2026};
    \item Only in the clumpy scenario does a significant fraction, $\approx 10-20\%$ of old, age $>10$ Gyr, $\alpha-$poor stars form in the disk (Section \ref{sec:old}, Figure \ref{fig:fraction});
    \item In the clumpy models, the $\alpha$-poor/thin disk emerges concomitant with the GSE merger, therefore it is not the GSE that induces its formation.
\end{itemize}
%
\section*{ACKNOWLEDGMENTS}

We thank the referee for helping to improve the clarity of this work.
JA thanks Jianhui Lian for his hospitality and fruitful discussions during his visit to Yunnan University. JA, GL, HP, and LB thank all those involved with the multi-institutional \textit{Milky Way BR} Group during weekly discussions.
JA \& CL acknowledge funding from the European Research Council (ERC) under the European Union’s Horizon 2020 research and innovation programme (grant agreement No. 852839).
JA, ZYL, CYC, and XJL are supported by the National Natural Science Foundation of China under grant Nos. 12233001, 12533004, by the National Key R\&D Program of China under grant No. 2024YFA1611602, by a Shanghai Natural Science Research Grant (24ZR1491200), by the ``111'' project of the Ministry of Education under grant No. B20019, and by the China Manned Space Program with grant Nos. CMS-CSST-2025-A08, CMS-CSST-2025-A09 and CMS-CSST-2025-A11.
CL also acknowledges funding from the Agence Nationale de la Recherche (ANR project ANR-24-CPJ1- 0160-01).
LBeS acknowledges support from CNPq (Brazil) through a research productivity fellowship, grant no. [304873/2025-0].
L.B. thanks the financial support by the São Paulo Research Foundation (FAPESP), Brasil (Proc. 2024/16510-2)
KF acknowledges financial support from the European Research Council under the ERC Starting Grant “GalFlow” (grant 101116226).
TK acknowledges support from the NSFC (Grant No. 12303013) and support from the China Postdoctoral Science Foundation (Grant No. 2023M732250).
The simulations in this paper were run at the High Performance Computing Facility of the University of Lancashire.
This work made use of the Gravity Supercomputer at the Department of Astronomy, Shanghai Jiao Tong University.

\bibliography{ref}{}
\bibliographystyle{aasjournal}



\appendix
\restartappendixnumbering

\section{Rotation curves }\label{app:rotcurv}
Figure \ref{fig:rotcurve} shows the rotational curve of the models presented in this work. They are calculated using the profile class of the {\tt pynbody} library \citep{pynbody}, which estimates the circular velocity, $v_c$, from the total gravitational force generated by the particles. Besides showing the rotational curve from the total mass (black), we show the contributions of DM (green), stars (blue), and gas (red).

\section{Orbit circularity of stars formed in the inner galaxy}\label{app:orbcirc}
As in Section~\ref{sec:chembi}, Figure~\ref{fig:ofe1dcirc} presents the [O/Fe] distribution for stars with $-0.7 < {\rm[Fe/H]} < -0.2$, $4 < R/{\rm kpc} < 12$, and $|z| < 3,{\rm kpc}$, now distinguishing stars formed in the inner galaxy, $R_{form} < 4,{\rm kpc}$, by their orbital circularity $\eta_s$. Stars on more circular orbits, $\eta_s > 0.9$, predominantly increased their angular momentum $L_z$ since formation while preserving orbital circularity, indicative of churning \citep{sellwood-binney2002}. These stars underwent radial migration, with their guiding radii shifting from formation to the present day. In contrast, stars on more radial orbits, $\eta_s < 0.9$, experienced blurring \citep{sellwood-binney2002} or rapid scattering during clump formation. These stars are temporarily present in the outer disk regions due to their eccentric orbits. \par
In the clumpy models, $\sim 62-74\%$ of stars formed in the inner galaxy ($R_{form} < 4,{\rm kpc}$) that contribute to the $\alpha-$rich disk component have dynamically hot orbits ($\eta_s < 0.9$). This fraction is reduced in non-clumpy models, though only the isolated non-clumpy model -- which lacks a chemical bimodality -- shows a significant difference.
\begin{figure*}
\centering
    
    \includegraphics[width=\linewidth]{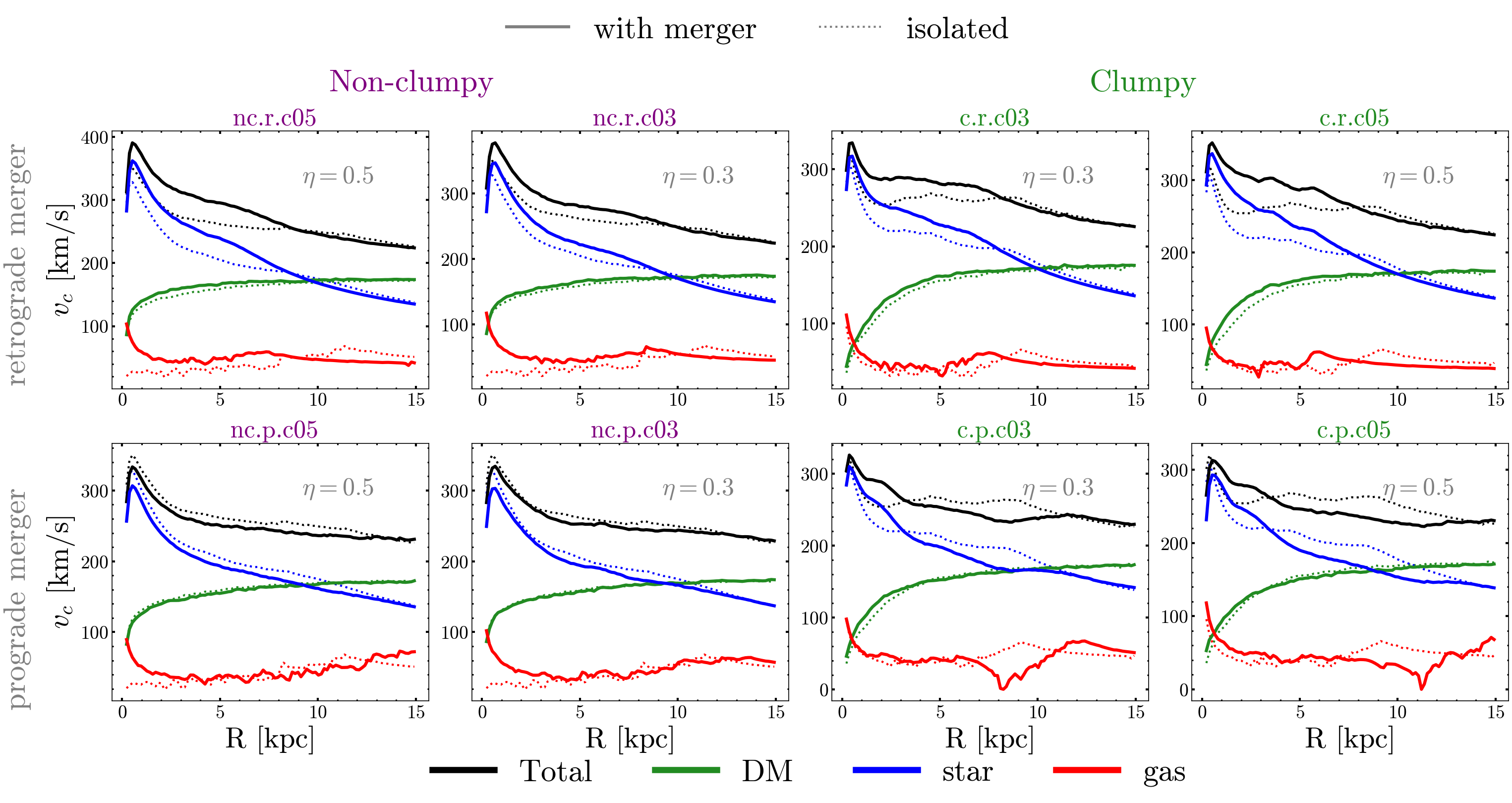}
        \caption{The rotation curve of the models used in this work. The dark matter, star, gas, and total mass contribution are shown in green, blue, red, and black solid lines. The dotted lines correspond to the isolated clumpy/non-clumpy models counterparts. }
    \label{fig:rotcurve}
\end{figure*}

\begin{figure*}
\centering

    \includegraphics[width=\linewidth]{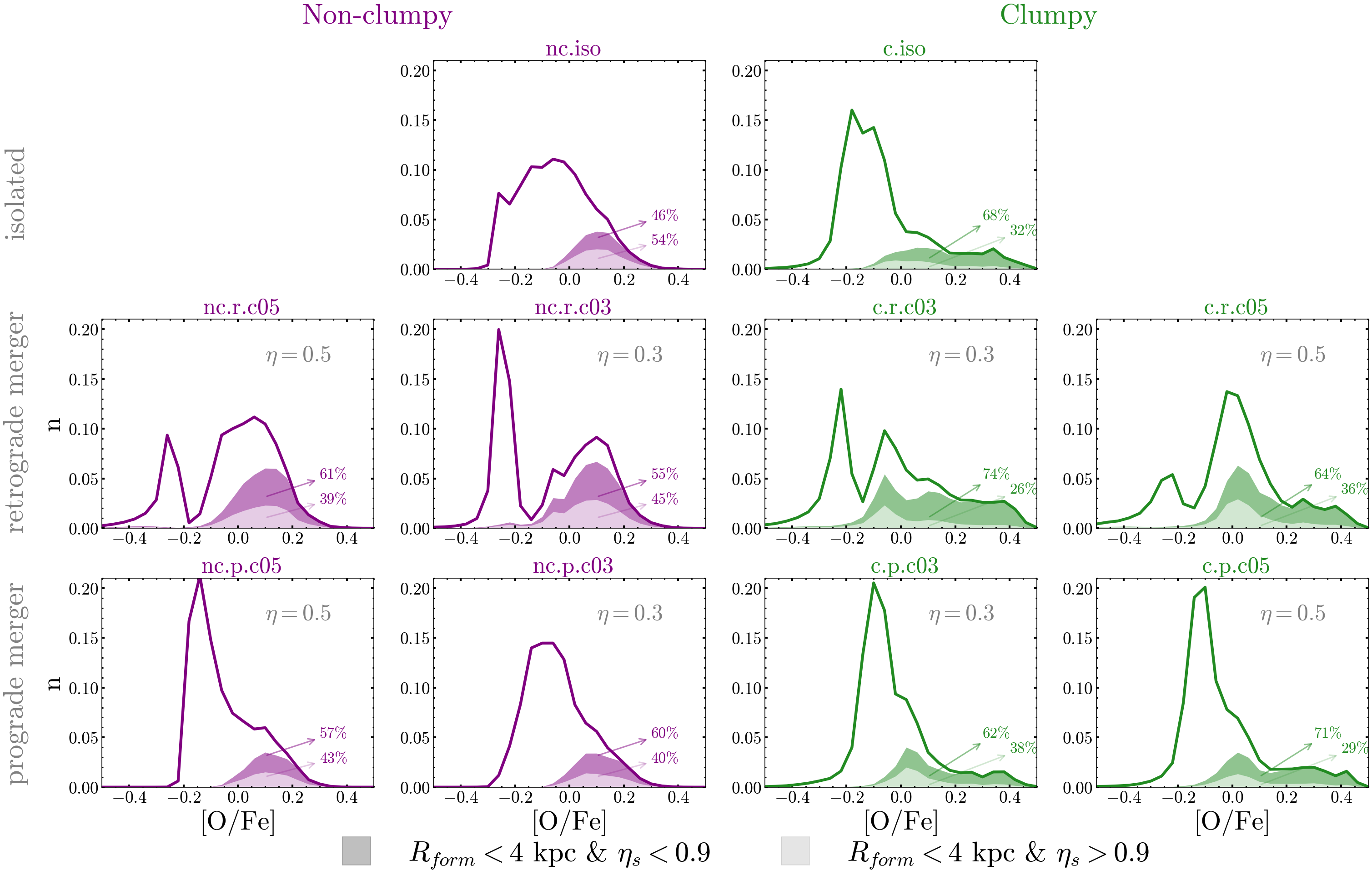}
    
    \caption{[O/Fe] histogram for the interval  $-0.7< {\rm [Fe/H]} < -0.2$, $4 < R/{\rm kpc} < 12$, and $|z|<3\, {\rm kpc}$. The dark and light shaded areas correspond to the contribution of stars born in the inner part of the galaxy, i.e. $R < 4$ kpc, with orbital circularity $\eta_s<0.9$ and $\eta_s>0.9$, respectively. Overall, from the stars born in the inner part of the galaxy that contribute to the $\alpha-$rich peak at the disk, there is a larger fraction with hotter orbits in the clumpy models. }
    \label{fig:ofe1dcirc}
\end{figure*}

\end{document}